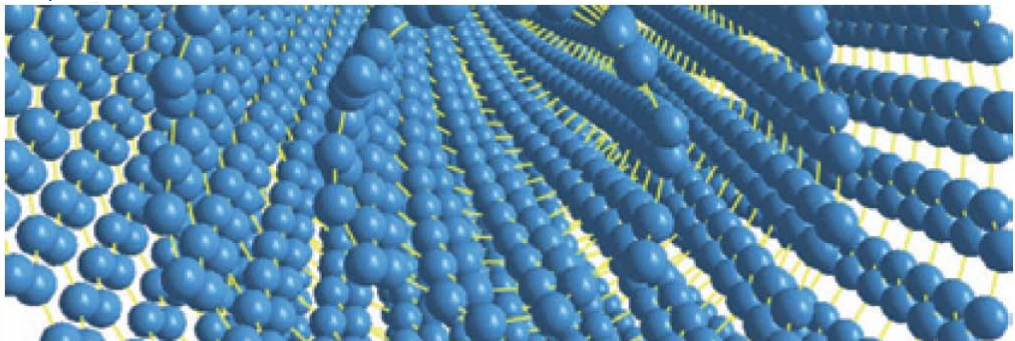

# NANO TREND

Observing nanoscience and technology worldwide  Newsletter n°2, February 2009

# Studying nano-districts gives a new view on nanoscience

**Data from 200 nano-districts – representing over 80% of articles worldwide – quantifies the shift towards Asia; at a national level several previous big players decline in significance.**

Research has established that nanotechnology production is concentrated at a regional level. We know that such districts encourage the flow of knowledge, including "tacit knowledge", between the various players (in particular universities and hi-tech companies) and that this knowledge-exchange favours the development of scientific activity. In addition, regional public policies have been seen to encourage geographic concentration in order to stimulate economic development, and virtually every state in the USA has a district-development strategy as part of its economic development plan.

This newsletter analyses the emergence of districts in nanotechnology and the factors affecting their growth in order to understand the drivers of economic development. By analyzing the top 200 nano-districts, Nanotrendchart (*www.nanotrendchart.org*) sheds light on the clusterization process: Does geographical location matter? How significant is size? Does the area of specialization play a role? What are the dynamics affecting growth? The results are intriguing: academic production is seen to be highly concentrated, with fifteen districts representing over half of published articles; Asian districts are clearly growing significantly faster than those elsewhere; and district growth in the domain of engineering and physics is more rapid than in other disciplines.

### Where are the nano-districts located?

In terms of large districts the Triad (North America, Japan and Europe) continues to play a central role, as can be seen in Table 1. Twelve districts represent 20% of published articles, with Asia representing about two-thirds of these. The replacement, since 1996, of two European districts (London and Berlin) by Shanghai and Singapore is evidence of Asia's growth. As a further example, since 2006 St Petersburg is no longer one of the 35 largest districts worldwide.

**Table 1: The top 35 nano-districts in 2005**

| District | Country | Rank | No. of articles | District | Country | Rank | No. of articles |
|---|---|---|---|---|---|---|---|
| Tokyo | Japan | 1 | 25,296 | New York | USA | 19 | 6177 |
| Beijing | China | 2 | 19,692 | Los Angeles | USA | 20 | 5973 |
| Kyoto | Japan | 3 | 16,827 | Madrid | Spain | 21 | 5970 |
| Seoul | South Korea | 4 | 13,529 | Hsinchu | Taiwan | 22 | 5921 |
| Berkeley | USA | 5 | 11,641 | Louvain | Belgium | 23 | 5782 |
| Paris | France | 6 | 11,550 | Hong Kong | China | 24 | 5665 |
| Tsukuba | Japan | 7 | 11,159 | Grenoble | France | 25 | 5519 |
| Shanghai | China | 8 | 9849 | Julich/Aachen | Germany | 26 | 5305 |
| Washington | USA | 9 | 9643 | Delft | Netherlands | 27 | 5264 |
| Moscow | Russia | 10 | 7911 | Nanjing | China | 28 | 4595 |
| Cambridge/Boston | USA | 11 | 7887 | St Petersburg | Russia | 29 | 4350 |
| Berlin | Germany | 12 | 7662 | Taipei | Taiwan | 30 | 4266 |
| London | England | 13 | 6720 | Cambridge | England | 31 | 4259 |
| Singapore | Singapore | 14 | 6650 | Changchun | China | 32 | 4222 |
| Nagoya | Japan | 15 | 6469 | Evanston/Chicago | USA | 33 | 4186 |
| Sendai | Japan | 16 | 6464 | Dresden | Germany | 34 | 4133 |
| Taejon | South Korea | 17 | 6457 | Mainz | Germany | 35 | 4116 |
| Zurich | Switzerland | 18 | 6284 | | | | |

In addition, our research gives an enhanced profile to countries previously absent from the global picture: Eastern Europe and several former USSR states feature as emerging districts. Three Polish districts, and Sofia, Minsk Ljubljana, Bratislava, Kiev, Bucharest, Prague, Budapest and Belgrade are all represented. Interestingly, although activity in nanoscience is concentrated in a limited number of areas, there is also diversification: several countries not traditionally associated with previous waves of development now feature in our global view. We can see new small districts in Latin America (Mexico and Buenos Aires) and in the Mediterranean area (Istanbul, Ankara, Cairo and Teheran).

## How big are the nano-districts?

As one might expect, academic production is highly concentrated, with more than 90% of articles generated by the USA and Canada, Asia and Europe. More interestingly, in total, 22 countries exceed the "1% of publications" (5500 articles) threshold. Within the European Union and associate countries, 10 countries including Israel and Switzerland, have achieved this level. What is surprising is that almost 40% of total scientific production worldwide takes place within only 35 districts and more than 70% of production in 200 districts.

Table 2 shows the number of districts of different sizes for each of the major geographic zones. Production of academic articles is clearly more geographically concentrated in Asia especially in South Korea and Japan where large and medium districts are more significant than in Europe and in the USA.

In line with the arrival of the new players described above, the number of small and emergent districts is considerably higher in the European zone than in Asia. Worldwide, clusterization remains highly concentrated in Europe; polarized in the USA (northeast and south-west); concentrated in Japan and South Korea; and dispersed in China.

## The dynamics of growth

Given the importance of districts for economic development, it is important to explain the factors affecting district evolution. Table 3 shows the average district growth-rate by geographic zone from 1998 to 2006.

Table 2: The number of nano-districts by area

| Geographic area | Number of districts | | | | Total | % of articles |
|---|---|---|---|---|---|---|
| | Large 10,000–40,000 | Medium 5000–10,000 | Small 2000–5000 | Emergent 1000–2000 | | |
| Europe | 1 | 9 | 40 | 31 | 81 | 34.4% |
| North America | 3 | 4 | 24 | 21 | 52 | 23.7% |
| Asia | 7 | 9 | 21 | 12 | 49 | 33.4% |
| Other | 1 | 1 | 7 | 9 | 18 | 8.5% |
| **Total** | **12** | **23** | **92** | **73** | **200** | **100.0%** |

Table 3 Nano-district growth by geographic

| Geographic area | Total |
|---|---|
| Europe | 10.2% |
| North America | 11.2% |
| Asia | 18.0% |
| Other | 11.3% |
| **Total** | **12.9%** |

Table 3 has several interesting features:

- The average annual growth rate of publications in nanotechnology is very high, around 13%, whereas the whole web of knowledge (Thomson Scientific ISI database) is growing at only about 3% yearly.

- Asian districts are growing significantly faster than those in the USA and Europe. The growth rate of American districts is 37% slower than those in Asia. However, within Asia there are higher variations in district growth than elsewhere. Two factors may account for this. Firstly, the mean size of Asian districts is smaller than those in the other zones; as smaller districts grow faster, Asian districts witness more rapid growth. Secondly, the largest Asian districts are growing fastest.

## How and where are nano-districts specializing?

For each geographical zone our data gives us an insight into the different specialization and its significance, and the importance of the zone. Table 4 summarizes the results.

Geographic specialization is highly heterogeneous. North America specializes in life sciences, with less emphasis on the physical and chemical sciences. On the other hand, Asia specializes in Engineering, computing and material sciences. The European profile is more balanced.

Table 4 suggests that:

- Asia is more specialized in Engineering, Computing and Technology, Electronics and Telecommunications and, to a lesser extent, in Physical, Chemical and Earth sciences which are also the faster-growing specialities.

- Asia has a tendency to specialize in all the life sciences which grow slower than other specialities. The USA and Canada are less specialized in Engineering, Computing and Material sciences and Electronics and Telecommunications – the latter following a global trend.

Table 4 Area of specialization for geographical zones

| | Districts | | | | |
|---|---|---|---|---|---|
| Specialization | Asia | Europe | North America | Other | Worldwide (% of articles) |
| Physical, Chemical & Earth Sciences | 105 | 101 | 88 | 111 | 51 |
| Engineering, Computing & Technology | 116 | 96 | 80 | 103 | 20 |
| Electronics & Telecommunications | 108 | 93 | 99 | 91 | 15 |
| Life Sciences & Clinical Medicine | 87 | 223 | 365 | 116 | 12 |
| **Percentage of articles per area** | **33** | **34** | **24** | **9** | **98** |

*Notes on Table 4:* We have used an index of 100 to indicate an average level of specialization; higher than 100 shows the area to be highly specialized; less than 100 means there is no specialization. The last column reveals the percentage of articles in each category. The total is less than 100% as marginal categories have been ignored.

- The European zone has an intermediate profile, with no particular specialization other than some emphasis on Engineering, Computing and Material sciences, and Electronics and Telecommunications.

## What influences the way nano-districts develop?

The rationale for nano-district growth is currently a topic of hot debate amongst policy makers and corporate strategists. Our results will be of interest to regional development agencies, corporate managers, public bodies and policymakers since they have clear implications for how economic growth is developped and sustained.

Our research has focussed on two different dimensions: structural variables which are determined by the environment (initial size, location, profile of specialization) and leverage variables (which shape the district in the short and medium term, for exemple the number of institutions involved, the openness of the district, its scientific diversity and the influence of public policy).

## How significant are structural variables?

We looked at a number of variables: size, geographic area, area of scientific specialization, fragmentation amongst different institutions (university laboratories, etc.), openness (the degree of collaboration between authors within and outside a district as measured by co-authorship amongst actors), geographical proxymity, and the influence of public policy.

Our results show that structural aspects explain 66% of the variation in growth, and that these play a central role in creating a favourable environment for innovation. Of particular interest is "path dependency" – where an existing activity acts as a catalyst for growth.

Nanotechnology firms tend to be located adjacent to large firms already involved in one of the parent disciplines, large universities carrying out research in nano-related technology, or large technology research establishments.

## Do leverage variables play a role?

Regarding this second category of variable we have several interesting findings:

1. Increased **scientific diversity**, through the development of specific scientific teams, increases the growth rate of the district. Institutional and scientific diversity go hand in hand, suggesting that organizational barriers within districts do not slow down information circulation.

2. Whereas scientific and institutional diversity impact positively on district growth, the **degree of openness** has a negative impact: higher openness reduces district growth. This may be a result of geographical patterns of collaboration, as seen in Asian districts, which grow faster and show a low proportion of co-authorship outside the district.

3. With the growth of the internet, globalization and modern communication technologies, companies are increasingly able to resource globally. One might then expect geographical location to play a diminishing role. Interestingly, however, the production of scientific articles appears to be increasingly concentrated. The richness and diversity of the local environment, and the closeness of the relationships amongst key players and the entrepreneurial spirit, are all key to district development. Given the importance of this intense scientific environment, **geographical proximity** is therefore considered a key factor in district growth. Howe-ver, it is not necessary the same places which are growing faster. If California and Massachusetts have been growing very fast during the biotech revolution, Asia is outperforming the North America for nanotechnologies.

4. The factors affecting district evolution are weakly by **public policies**. Specialization may be enhanced by the development of national or European programme, e.g. the *Pole d'excellence* policy in France.

## What do our results mean for policymakers?

The results confirm firstly that the trend towards Asia is very strong and that the initial advantage of the triadic countries is rapidly vanishing in term of the production of articles. This is less pronounced in terms of highly cited articles as the most cited articles are mainly produced by USA- and Europe-based scientists.

A second major conclusion underlines the high concentration of global research, with only 200 districts, and 35 large ones, worldwide. This is not simply a result of previous agglomeration trends since – as we have seen – it is Asia that now appears as a major player world science, challenging USA domination. Even in Europe, geographical distribution is shifting significantly. For instance, Cambridge is no longer in the top 35 districts.

Only a few large districts worldwide are of sufficient size to benefit from the growth associated with a diverse knowledge base. As emergent, small- and medium-size districts remain specialized, we would therefore recommend that National and regional public policies be integrated to support their development. In this way smaller, or specialized districts, can create strong ties with fast-growing areas to maintain their growth.

---

**Box 1 Identification and construction of 'nanodistricts': methodology**

Data collection was based on publications systematically collected from the ISI/web of Science through a specific query based on keywords. The database counts around 600,000 publications in nanotechnology from 1998 to 2005. We define "participation to publication" as the participation from one institution to a publication. The number of participations is higher than the number of publications, with 1,055,130 participations.

The software developed for geolocalizing addresses was based on a semantic recognition of cities (and states, regions or prefectures for Federal countries). We used the geolocalization engine, MapPoint (2006), to allocate cities their data in the World Geodesic System of 1984. The treatment is automatic (residual cases are also proposed for manual handling). This enabled to 94% of all addresses (just under 1 million) to be geolocalized and 97% of total articles. All cities with 1000 addresses and more were then identified.

Nanodistricts were constructed using a geographic rule: all addresses were within a 50km radius. This applied worldwide except for Japan, Korea and Taiwan where the radius was 30km. When two districts had more than 20% of addresses in common (calculated on the smaller district) they were merged.

**Interview with Nicolas Leterrier, Chief Representative at Minalogic, Grenoble**

Q: *Minalogic is a well known centre of excellence in nano-electronics and nano-technology. From your perspective do these results surprise you in any way?*
The districts that are listed in the 'Top 35' table are not those that one would normally expect to see. For example, the top 6 districts do not have any manufacturing activity. This includes Paris – which features by virtue of the large number of academic institutions who are able to publish freely because there is no problem with IP protection for manufacturing. On the other hand, a large manufacturing centre such as Taipei is ranked relatively low. I'm surprised to see Russia placed so high. It could be because of the legacy of its military activities. In the USA it's no surprise to see Berkeley's ranking – with its connection with INTEL. Ditto New York, LA and Chicago, with its links to the car industry. So yes, familiar faces. It's just Washington that's curious.

Q: *What about the areas of specialization – for example, in your view, what explains Asia's relatively low representation in the life sciences?*
You also need to be aware of the differing product-development life cycles. This is much more extended for life sciences – around 10 years – whereas the other scientific specializations it's nearer two or three years. With regard to the life sciences the fact that, in Asia, the approach to medicine tends to be preventative, and based on traditional values rather than high-tech ones, could explain the relatively low emphasis given to this discipline. By contrast, in Europe the presence of the big pharmaceutical companies is a ready reason for the high level of activity in this area.

Q: *What about the factors affecting growth?*
I would agree that geographic proximity, scientific diversity and the degree of openness are all key factors. But, just to give an example of an exception to the rule, New York state has become a specialist in nano-electronics because of the commitment to making it work. New York's growth rate has been exponential, as a result of a blend of applied research, and industrial and academic research. And that was starting from virtually nothing. On the other hand, in Asia development has tended to be based around large cities that are open to exterior influence, in the knowledge that this will attract students.

Q: *And the different growth rates?*
It's no surprise to see Asia leading, fuelled by its strong desire to be independent. In China they have a process whereby existing patents are studied and the science around them recreated and improved. Even though the technology may be obsolete it's regarded as a method for learning. If the patent is registered in, say, the US patent office, the product can be re-launched in another area where the patent isn't protected.

Q: *How does Grenoble's profile as a nano-centre differ from other nano-districts?*
With the decline of Dresden there is no competition in Europe and Grenoble is now the number one nano-electronics centre, in Europe, and is recognized as being in the top 4 globally (the other three being New York state, Hsinchu (Taiwan) and Tsukuba (Japan). Grenoble's competitive advantages lie in having mastered three technologies: low power consumption, radio frequency communication and imaging. Globally we could become the world champions of low power consumption and radio frequency communication – and leave high power to the Americans. I can see the potential for gaining market share by being smart: developing multi-faceted products using low consumption, miniaturization and communications knowledge. Our skill in these areas was our strength when working with Nokia and Schneider. Apart from low consumption the next two market niches to emerge will be health and energy. The 'Grenoble clover leaf' is based on these two markets, along with media and communications. These will be the fundamental needs of the future.

Q: *And finally… are our results on the different sizes of nano-district as you would expect?*
The issue here is that, in Europe, we 'hedge our bets', investing a bit here and there rather than concentrating it on a potential leader. It's linked with European politics and notions of political correctness which encourage multi-national partnerships rather than focusing on excellence. Frustrating but that's the world that we live in!

**Interview by Corine Genet,
Associate Professor at Grenoble Ecole de Management**

ACKNOLEDGMENT: This work is supported by French National Agency (ANR) through the Nanoscience and Nanotechnology Program (Project NANOBENCH n°ANR-07-NANO-026).

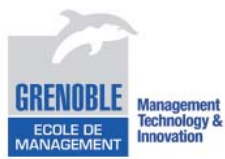
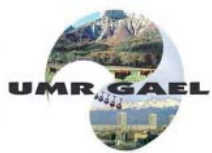